\documentclass[12pt,side,epsfig,citesort,color]{article}
\usepackage{booktabs}
\usepackage{setspace}
\usepackage{color}
\usepackage{geometry}
\usepackage{epsfig}
\usepackage{amssymb}
\usepackage{amsmath}
\usepackage{amsfonts}
\usepackage{graphicx}
\usepackage{physics}
\usepackage{bm}
\usepackage{epsfig}
\usepackage{cite}
\usepackage{subfigure}
\usepackage{booktabs}
\usepackage{caption} 
\begin{document}
\voffset=0.9cm

\title{                                                               
\vspace*{-3.2cm}
Exploring the attosecond laser-driven electron dynamics in the hydrogen molecule with different real-time time-dependent configuration interaction approaches}
\author{%
Aleksander P. Wo\'zniak\textsuperscript{1} \thanks{Corresponding author:                          
e-mail address: ap.wozniak@uw.edu.pl}, 
  Maciej Lewenstein\textsuperscript{2,3}
  Robert Moszy\'nski\textsuperscript{1}      \\                                    
{\sl \textsuperscript{1}Faculty of Chemistry} \\
{\sl University of Warsaw}\\                             
{\sl Pasteura 1, 02-093 Warsaw, Poland}\\          
{\sl \textsuperscript{2}ICFO - Institut de Ciencies Fotoniques} \\ 
{\sl The Barcelona Institute of Science and Technology,}\\
{\sl Av. Carl Friedrich Gauss 3, 08860 Castelldefels (Barcelona), Spain}\\          
{\sl \textsuperscript{3}ICREA} \\                                       
{\sl Pg. Llu\'is Companys 23, 08010 Barcelona, Spain}\\          
}
\maketitle

\begin{abstract}
Time-dependent quantum chemical methods coupled to Gaussian basis sets are gaining popularity in modeling the electron dynamics of atoms and molecules interacting with intense laser fields. Two approaches most widely used for this purpose, the real-time time-dependent configuration interaction singles and the real-time time-dependent density functional theory, both have their limitations, so the development of more accurate yet computationally efficient time-dependent methods is still in demand. In this work we explore the applicability of the real-time time-dependent configuration interaction singles and doubles (RT-TDCISD) in modeling strong field phenomena. Since the main drawback of RT-TDCISD is its unfavourable scaling, we develop several algorithms for reducing the effective propagation space by selecting these CISD eigenstates that should have dominant contribution to the time-evolution of the wavefunction. We test them by performing calculations of the high harmonic spectra of the H\textsubscript{2} molecule. We find out that the laser-driven electron dynamics is mostly realized in a very small subspace of eigenstates dominated by single excitations, that constitutes about one percent of the whole CISD eigenspectrum. Therefore, by properly selecting this subspace one can reduce the dimension of the propagation equation by two orders of magnitude without affecting the time-resolved observables.
\end{abstract}

\begin{center}
KEYWORDS
\end{center}

\noindent Laser-driven electron dynamics, nonlinear optics, high harmonic generation, time-dependent Schrödinger equation, real-time time-dependent configuration interaction

\tableofcontents

\section{Introduction}

With the rapid development in attophysics, the real-time time-dependent quantum chemical methods coupled to specifically designed Gaussian basis sets gain increasing interest in solving the time-dependent Schrödinger equation, a task required for the description of electron dynamics during attosecond processes \cite{ishikawa2015,goings2017,coccia2022}.
Among various computational chemistry approaches, the two that are currently most widely used in theoretical attoscience are the real-time time-dependent density functional theory (RT-TDDFT) \cite{runge1984,tong1998,castro2004} and the real-time time-dependent configuration interaction restricted to single excitations (RT-TDCIS) \cite{klamroth2003,huber2005,rohringer2006,krause2007,schlegel2007}.
Both of them have been successfully applied to a wide range of systems from simplest one- or two-electron atoms \cite{rohringer2006,luppi2013,castro2015,coccia2016a,pabst2016,wozniak2021,wozniak2022} to complex biological molecules \cite{lopata2011,luppi2021,morassut2022} and nanostructures \cite{kuisma2015,wopperer2015,rossi2017,mrudul2020}.
Their popularity relies chiefly on great scalability with the number of electrons, allowing for propagating large systems for timescales of femto- or even picoseconds.
However, both of them also have their shortcomings.

RT-TDCIS is a well established wavefunction theory, in which the wavefunction of the examined system is expanded in the basis of $N$-electronic ground and excited states.
By employing only single excitations it offers a very favourable linear scaling with the number of occupied spinorbitals.
However, being a counterpart of the HF method for excited states, it completely neglects any dynamical correlation effects.
Each year, more experimental evidence is being reported for novel attosecond phenomena, such as the giant dipole resonances \cite{barillot2015,zhong2020} or the Cooper minima in photoionization cross sections \cite{bertrand2012,schoun2014,alexandridi2021}, the theoretical explanations of which often rely on dynamical electron-electron interactions.
It therefore becomes inevitable that when trying to model such subtle effects, one has to eventually abandon RT-TDCIS in favour of correlated methods.

By contrast, electronic correlation is easily included in RT-TDDFT by using an approximate and appropriate exchange-correlation potential of choice. As opposed to the wavefunction theories, RT-TDDFT employs propagation Kohn-Sham orbitals within a single determinant, which usually provides efficiency comparable to or even better than that of RT-TDCIS.
Among the disadvantages of RT-TDDFT one can mention the non-linearity of the equations of motion for the KS orbitals \cite{bedurke2021}, and the inability to describe open-shell singlet states, which is a common feature of all single-determinant approaches \cite{isborn2009,sato2014}.
The latter has some severe consequences for the description of strong field processes, such as the high-harmonic generation (HHG), in systems with a closed-shell ground state.
HHG, one of the key aspects of attophysics, is a highly nonlinear optical process occuring in atoms and molecules placed in laser fields of extreme intensity and relatively long wavelengths, in which the frequency of the exciting field is converted into a series of its integer multiples \cite{antoine1996}.
As explained by celabrated analytical models of HHG, such as the three-step model (3SM) \cite{krause1992,schafer1993,corkum1993} or Lewenstein model \cite{lewenstein1994}, the key component of the HHG process is the field-induced tunelling ionization of a single electron, followed by its reacceleration and recombination with the residual ion accompanied by the emission of light.
Therefore, the correct description of HHG in closed-shell systems requires involving transitions to open-shell singlet states that are accesible only within the wavefunction (multideterminant) framework.

The above-mentioned limitations of currently used real-time propagation methods entail the development of novel, more precise theoretical approaches that will combine the strengths and avoid the weaknesses of RT-TDCIS and RT-TDDFT, while at the same time ensuring similar computational efficiency.
One of the promising candidates is the real-time time-dependent equation-of-motion coupled-cluster (RT-TDEOMCC) method, however its current implementations are troubled by numerical instabilities \cite{kvaal2012}.

In our recent paper we presented proof-of-concept RT-TDCISD (effectively RT-TDFCI) calculations of HHG spectra in the helium atom \cite{wozniak2022}. We used Gaussian basis sets sufficiently large to describe the evolution of the wavefunction at laser intensities above $10^{14}$ W/cm\textsuperscript{2}, frequently used in attosecond experiments.
It was shown that RT-TDCISD provides improvements in terms of features of of the spectra the spectra such as position of harmonic cutoff or intensity of harmonic peaks.
However, the scale of these corrections is in general not commensurate with the huge increase in the computational costs related to including double excitations in the CI wavefunction.
This issue is exacerbated when increasing the laser intensity, as higher field's ponderomotive force makes the ionized electron travel longer distances.
This, in turn, requires using larger, more diffuse basis set, making RT-TDCISD unfeasible from a certain point on.
Therefore, in this paper we explore possible ways of decreasing the computational cost of RT-TDCISD while at the same time keeping it able to capture the essential correlation effects in laser-induced electron dynamics.
We introduce and implement several algorithms of decreasing the CISD space the real-time propagation is performed in.
They rely either on truncating the subspace of virtual orbitals from which the configurations are constructed, or on selecting certain CI states after solving the time-independent CI problem.

We evaluate the performance of the proposed algorithms by calculating the HHG spectra of the hydrogen molecule, as the example of a multicenter, many-electron system. The diatomic molecules have been in the center of attoscience for the last few decades, being studied both experimentally \cite{schmidt1994,cornaggia1994,lynga1996,lappas2000,velotta2001,hay2002,lein2002a} and theoretically, often using approximate models \cite{lappas2000,ivanov1993,zuo1993,yu1995,moreno1997,kopold1998,bandrauk1999,averbukh2001,kreibich2001,suarez2016,suarez2017}. The molecular HHG offers additional degrees of freedom and promising possibilities, such as the alignment of the molecular axis with respect to the laser field polarization axis. Specifically for a diatomic system, the existence of a distinctive quantum-interference minima pattern in the spectra and its dependence with the molecular orientation have been theoretically predicted \cite{lein2002b,lein2003,ciappina2007}. This pattern is due to a destructive interference from the high-harmonic emission at spatially separated centers; it allows to estimate the internuclear distance from the HHG spectra. In addition, the chance of controlling the phases and improving the phase-matching condition opens novel routes of investigation. The distinctive features of the molecular HHG spectra can be used to retrieve structural and dynamical information in simple molecules \cite{itatani2004}. Finally, from the molecular HHG spectra the temporal evolution of the electronic wavefunction can be directly recovered \cite{smirnova2009,haessler2010,kraus2015}.  

The paper is constructed as follows. In Section II we present the outline of the RT-TDCI theory and the details of proposed algorithms, as well as provide computational parameters of the simulations. In Section III we present and discuss the numerical results. Finally, in Section IV we conclude our work.

\section{Theoretical methods}
\subsection{The RT-TDCI theory}
We aim at solving the time-dependent Schrödinger equation (TDSE),
\begin{equation} \label{tdse}
i \frac{\partial}{\partial t}\Psi(t) = \hat{H}(t)\Psi(t),
\end{equation}
for the electronic wavefunction of the hydrogen molecule subject to a pulse of intense laser radiation, in a non-pertubative manner, using time discretization. The Hartree atomic units are used throughout the paper. The time-dependent Hamiltonian $\hat{H}(t)$ consists of the ground state H\textsubscript{2} Hamiltonian and the field interaction operator represented in the length gauge and in the dipole approximation:
\begin{equation}
\hat{H}(t) = \hat{H}_0 - \hat{\mu}\mathcal{E}(t).
\end{equation}
Here, $\hat{\mu}$ is the molecular dipole operator and $\mathcal{E}(t)$ is the time-dependent electric component of the external laser field.
The time dependent wavefunction in RT-TDCI is written as a linear combination of $\hat{H}_0$ eigenstates with time-dependent coefficients:
\begin{equation}
\Psi(t) = \sum_k C_k(t) \Psi^\textrm{CI}_k,
\end{equation}
where $k$ is the electronic (ground or excited) state index.
Each CI eigenstate is further expanded in the basis of the reference Slater determinant - the solution of the restricted Hartree-Fock (RHF) equations - and the $n$-tuply excited Slater determinants,
\begin{equation}
\Psi^\textrm{CI}_k = c_{0,k} \Phi^\textrm{RHF} + \sum_{i}^{occ.} \sum_{a}^{virt.} c_{i,k}^a \Phi_i^a + \sum_{i,j}^{occ.} \sum_{a,b}^{virt.} c_{ij,k}^{ab} \Phi_{ij}^{a,b} + ...
\end{equation}
The maximum excitation number defines the CI (and conseqeuently also the RT-TDCI) level of theory. In our case of the H\textsubscript{2} molecule there are only two electrons, so the CISD level is equivalent to full CI. Since we are considering a closed-shell system and the time-dependent Hamiltonian does not contain any spin-dependent terms, the time-evolution of the wavefunction is limited to singlet-state manifold, and therefore in actual calculations we use singlet spin-adapted configuration state functions (CSFs or simply configurations) rather than Slater determinants. The details of the construction of singly- and doubly-excited CSFs can be found in the literature \cite{szabo2012,krause2007,wozniak2022}.

The coefficients $c_{0,k}$, $c_{i,k}^a$ and $c_{ij,k}^{ab}$ are obtained by solving the time-independent CI eigenequation,
\begin{equation} \label{ci}
\mathbf{H}_0^\textrm{CI}\mathbf{c}_k = E^\textrm{CI}_k \mathbf{c}_k.
\end{equation}
Afterwards, the CI vector of coefficients $C_k(t)$ is propagated in time starting from the obtained ground state. In this work Eq. \eqref{tdse} is approximated by the Crank-Nicolson propagator:
\begin{equation} \label{crank}
    \left(\mathbf{I}+\frac{i\Delta t}{2}\mathbf{H^\textrm{CI}}(t+\Delta t/2)\right)\mathbf{C}(t+\Delta t) = \left(\mathbf{I}-\frac{i\Delta t}{2} \mathbf{H^\textrm{CI}}(t+\Delta t/2)\right)\mathbf{C}(t),
\end{equation}
where $\mathbf{H^\textrm{CI}}_{lk} = \delta_{lk} E^\textrm{CI}_k - \mathcal{E}(t)\bra{\Psi_l}\hat{\mu}\ket{\Psi_k}$.

After the real-time propagation is finished, the time-resolved observables are retrieved from $C(t)$ and the HHG spectrum is calculated as the squared modulus of the Fourier transform of the time-resolved molecular dipole moment, normalized over the total simulation time $[t_i,t_f]$:

\begin{equation}
I_{\mathrm{HHG}}(\omega) = \abs{ \sum_{k,l}\bra{\Psi_l}\hat{\mu}\ket{\Psi_k}\frac{1}{t_f-t_i}\int_{t_i}^{t_f}C_l^*(t)C_k(t) e^{i\omega t}dt}^2.
\end{equation}

\subsection{Reducing RT-TDCISD propagation space}
In the most conventional realisation of RT-TDCI (which we will from now on refer to as ``standard'') Eq. \eqref{ci} is solved in the basis of all configurations possible at a given level of theory, constructed using the whole set of molecular orbitals (MOs). Usually the full diagonalization of the Hamiltonian matrix is performed, and the time-dependent CI vector is expanded in all obtained states. This is obviously the source of the much higher cost of RT-TDCISD when compared to RT-TDCIS: for instance, from $o$ occupied MOs and $v$ virtual MOs one can construct $\frac{1}{2}(o^2v^2+3ov+2)$) singlet CISD configurations and only $ov +1$ singlet CIS configurations. In modeling phenomena like HHG with discrete basis sets, a high density of states, especially those near and above the ionization threshold, is desirable, because it allows to accurately approximate the electronic continuum spectrum involved in the ionization process. However, one can expect that not every state has an equal contribution to the time-evolution and some of them may be in fact not needed, eg. simply because they are not energetically accessible under applied laser conditions. In this section we will discuss methods of identifying and \textit{a priori} excluding such redundant states and/or configurations, in order to speed up the real-time propagation.

Let us start by recalling the famous three-step HHG cutoff formula \cite{krause1992,schafer1993,corkum1993},
\begin{equation}
    E_{\mathrm{cut}} = I_p + 3.17 U_p = I_p + 3.17 \frac{\mathcal{E}_0}{4\omega_0}
\end{equation}
which predicts the energy of the highest harmonic generated by an atomic or molecular species with ionization potential $I_p$ in an oscillating electric field characterized by the ponderomotive force $U_p$. Here, $\mathcal{E}_0$ and $\omega_0$ are the field's amplitude and angular frequency, respectively. In this paper we approximate $I_p$ in the spirit of Koopmans' theorem, as $-\epsilon_{\mathrm{HOMO}}$.

In the simplest one-electron picture, $3.17 U_p$ is at the same time the energy of the highest virtual orbital to which an electron can be excited during harmonic generation. This is due to energy conservation, since the three-step model assumes that a single HHG event starts from and leads to the electronic ground state. Therefore, we can limit the configuration basis by constructing CSFs using virtual orbitals with $\epsilon < 3.17 U_p$ only or, more generally, introduce an auxiliary parameter $\lambda$ that allows us to adjust the threshold for maximum orbital energy, $\epsilon_{\mathrm{thr}} = 3.17 \lambda U_p$. In this way, setting e.g. $\lambda = 2$ means that in the calculations we use CSFs constructed only from orbitals with energies below twice the electron ponderomotive energy. Reducing the number of configurations lowers the computational cost of both the time-independent and the time-dependent problem, as well as the amount of memory required for storing the operator matrices in the configuration representation and in the CI representation. This procedure of selecting a subset of active orbitals is very well known from methods like restricted-active-space CI; in this work we just make the active virtual-orbital space dependent on the laser conditions. 

Shifting from the single-electron picture to the more accurate $N$-electron state picture, we can propose an alternate reasoning for truncating the space of CISD eigeinstates after we obtain them using the whole set of canonical MOs.
In this case, the highest continuum state able to emit harmonic radiation according to the three-step model has the excitation energy of $I_p + 3.17 U_p$.
Similarly to the truncation of the virtual-orbital space, we give some additional flexibility to this boundary, and introduce another auxiliary parameter $\eta$ that regulates the energy threshold for the CISD eigenstates to be included in RT-TDCISD propagation, $E_{\mathrm{thr}} = I_p + 3.17 \eta U_p$.
Just like the $\lambda$ parameter defined the subset of active MOs, the $\eta$ parameter analogously defines the subspace of active CI states, and setting either of the parameters to 1 corresponds to the respective 3SM prediction.
Like the previous one, this algorithm also allows for reducing the dimension of the CI representation, and thus accelerates the real-time propagation.
The dimension of the configuration representation remains unchanged in this case and the full CSF Hamiltonian matrix must be constructed.
However, it is still possible to speed up the solution of the time-independent CI equation by employing algorithms for partial matrix diagonalization, available eg. in the implementations of LAPACK numerical library.

In the last algorithm for reducing the CISD propagation space we also expand the time-dependent CI vector into a certain subset of CISD eigenstates.
This time, however, we will adopt a different criterion for choosing these eigenstates, which is no longer related to the excitation energies.
Numerous examples of good agreement between experimental results and theoretical predictions from 3SM and related approaches, such as the strong field approximation or single-active electron model, prove that in a wide variety of atoms and molecules HHG is indeed dominated by single-electron transitions \cite{amini2019}.
We reached a similar conclusion when examining relatively small differences between RT-TDCIS and RT-TDCISD spectra in helium \cite{wozniak2022}.
Therefore, we can deduce that the CISD states crucial for the laser-driven dynamics are mostly dominated by single excitations, and the dynamical correlation between the electron being ionized and the one within the residual ion is provided by a lesser admixture of double excitations.
On the other hand, it is well known that from a CISD calculation one obtains a huge number of states dominated by doubly excited configurations, that are not necessarily required for the description of HHG.
They may be useful for simulating other non-linear processes, such as non-sequential double ionization, that take place under different laser regimes.
However, for the purpose of HHG modeling, we can exclude these states from the RT-TDCISD space and perform the propagation using only states with a certain minimal contribution of single excitations.
To do this, after computing a full CISD eigenspectrum, for each state vector we calculate its squared partial norm as a sum of the squares of the coefficients $c_k$ corresponding to the reference configuration and to the singly excited configurations:
\begin{equation}
    N_k^{\mathrm{RS}} = \abs{c_{0,k}}^2 + \sum_i^{occ.} \sum_a^{virt.} \abs{c_{i,k}^a}^2
\end{equation}
and select only states with $N_k^{\mathrm{RS}}$ above some threshold.
This approach is conceptually similar to RT-TDCIS(D), which was also proposed in the literature as a method for including electronic correlation in the description of HHG \cite{krause2005,krause2007,schlegel2007}.
It is an extension of RT-TDCIS, in which the effects of double excitations are treated via perturbative corrections to CIS eigenenergies.
However, unlike RT-TDCIS(D), the number of states selected by our algorithm is not limited by the dimension of the CIS space.
We can vary the threshold for $N_k^{\mathrm{RS}}$ to obtain the distribution and density of states that ensures a proper balance between accuracy and computational efficiency.
The main drawback of the algorithm is that it requires a full diagonalization of the standard CISD Hamiltonian (unless of course it is combined with another truncation method) and thus is only able to reduce the computational costs of the real-time propagation.

The algorithm of selecting only single-excitation dominated states may on first sight look very efficient, since the fact that the number of singly excited configurations is smaller by orders of magnitude than the number of doubly excited configurations usually implies a similar tendency in the numbers of states dominated by single and double excitations.
However, it must be noted that not all double-excitation dominated states stand in contradiction with the single ionization assumption of 3SM.
In particular, the contribution of configurations in which one electron is excited to a continuum orbital and the other is excited to a bound orbital (we call them bound+continuum excitations) may have pivotal role in dynamic correlation, since they are responsible for the polarization of the electronic density in the residual ion caused by the ionized electron.
Discarding states with a predominant contribution of such configurations may be too excessive, and therefore we propose also a second variant of the algorithm, in which we calculate modified squared partial norms of CI state vectors,
\begin{equation}
     N_k^{\mathrm{RSBC}} = \abs{c_{0,k}}^2 + \sum_i^{occ.} \sum_a^{virt.} \abs{c_{i,k}^a}^2 + \sum_{i,j}^{occ.} \sum_a^{virt.} \sum_b^{bv.} \abs{c_{ij,k}^{ab}}^2
\end{equation}
where the index $bv.$ refers to summing over bound virtual orbitals (with energies $\epsilon_b < 0$). In this way, we exclude only states dominated by double excitations to two continuum orbitals.

\subsection{Computational details}
To assess the performance of the proposed algorithms, we solve the time discretized TDSE \eqref{tdse} for the hydrogen molecule in the fixed-nuclei approximation, at the equilibrium bond distance of 1.4 a.u. The laser pulse is represented by an oscillating electric field with a sin-squared envelope,
\begin{equation} \label{efield}
\mathcal{E}(t) = 
\begin{cases}
\mathcal{E}_0 \sin(\omega_0 t) \sin^2(\omega_0 t/2 n_c)&\text{if}\; 0 \leq t \leq 2\pi n_c/\omega_0, \\
0& \text{otherwise,}
\end{cases}
\end{equation}
where we use $\omega_0$ = 1.55 eV (corresponding to a wavelength of 800 nm) and the number of optical cycles $n_c$ = 20, so that the total duration of the pulse is approximately 2206 a.u. (53.4 fs). The polarization of the field is assumed perpendicular to the bond axis. We carry out calculations for two laser intensities: $I_0$ = $7 \times 10^{13}$ W/cm\textsuperscript{2} and $I_0$ = $10^{14}$ W/cm\textsuperscript{2}. The time-dependent CI vector is propagated with a timestep $\Delta t$ = 0.01 a.u., a value that provided stability of the calculations. After the pulse ends, we propagate the wavefunction for ten additional optical cycles, to ensure that the system reaches a stationary point.

The set of canonical MOs is obtained through the solution of the RHF equations in a Gaussian basis set adjusted specifically to describe highly excited and continuum electronic states.
We use Dunning's correlation-consistent aug-cc-pVTZ basis set, which we further augment with the so-called active-range-optimized (ARO) functions we introduced and tested in our previous works \cite{wozniak2021,wozniak2022}.
In this paper we compare two types of ARO augmentation.
In the first one we place a set of ARO functions at each atomic center (atomic augmentation), while in the second we place only one set of ARO functions in the center of charge of the examined system, in our case in the middle of the H\textsubscript{2} bond (midbond augmentation).
The latter augmentation scheme using different kinds of diffuse functions has already been used in HHG calculations of larger molecules \cite{bedurke2019}, and it is appealing due to a number of reasons.
First, introducing a new basis set center allows for obtaining states with higher angular momenta due to mixing of spherical harmonics at different centers.
This feature is highly desired in simulating laser-driven dynamics, when absorption of multiple photons is involved.
Second, midbond augmentation allows to use more Gaussian exponents compared to atomic augmention, when the total number of basis functions is kept the same.
Finally, an ionized electron at a sufficiently large distance from the molecule experiences the potential of the residual ion approximately as a point charge, so it's behavior should be reasonably described using single-center orbitals, analogously to Coulomb wave functions of a hydrogen-like ion.
In our calculations we use ARO functions with angular momenta from $l$=0 ($s$-type orbitals) to $l$=3 ($f$-type orbitals), fitted to set Slater orbitals with exponent $\zeta$=1.0 and principal quantum numbers from 1 to 100, according to procedure described in ref. \cite{wozniak2021}.
In the atomic augmentation scheme we place three ARO functions per angular momentum at each of the two atom, while in the midbond augmentation scheme we use a single ``ghost'' atom with six ARO functions per angular momentum, so that the total numbers of functions in both basis sets are equal.

Because the Gaussian basis sets we use are far from completeness, we have to account for the ionization losses during HHG. We employ the so-called heuristic finite lifetime model reported by Klinkusch \textit{et al.} \cite{klinkusch2009a}. Its main idea is to add imaginary ionization rates to energies of CI eigenstates beyond the ionization threshold:
\begin{equation}
    E_k^{\textrm{CI}} \rightarrow E_k^{\textrm{CI}} - \frac{i}{2} \Gamma_k^{\textrm{CI}} \quad \text{for} \; E_k^{\textrm{CI}} \geq E_0^{\textrm{CI}} + I_p,
\end{equation}
so that these eigenstates are interpreted as non-stationary. After this modification the time-dependent Hamiltonian matrix is no longer Hermitian, so the time-evolution becomes non-unitary. The ionization rates $\Gamma_k$ are calculated according to ref. \cite{klinkusch2009a}, as sums of single-orbital ionization rates $\gamma_a$,
\begin{equation}
\gamma_a = \theta(\epsilon_a) \sqrt{2\epsilon_a}/d_a,
\end{equation}
where $\theta(x)$ is the Heaviside step function and $d_a$ is a characteristic escape length of an electron on orbital with energy $\epsilon_a$. The expressions for $\Gamma_k^{\textrm{CI}}$ for the CIS and CISD level are, respectively,
\begin{eqnarray}
\Gamma_k^{\textrm{CIS}} = &&\sum_{i}^{occ.} \sum_{a}^{vir.} \abs{c^a_{i,k}}^2 \gamma_a, \\
\Gamma_k^{\textrm{CISD}} = &&\sum_{i}^{occ.} \sum_{a}^{vir.} \abs{c^a_{i,k}}^2 \gamma_a + \sum_{i,j}^{occ.} \sum_{a,b}^{vir.} \abs{c^{ab}_{ij,k}}^2 (\gamma_a + \gamma_b).
\end{eqnarray}
In our calculations we use a modification of the finite lifetime model introduced in our previous work \cite{wozniak2022}, in which we divide the virtual MOs into two groups and assign a different value of escape lenght $d$ to each group. For orbitals with energies $\epsilon_a < 3.17 U_p$ we set $d_1$ equal to the semiclassical electron quiver amplitude in an electromagnetic field, $\mathcal{E}_0/\omega_0^2$, while for orbitals with energies exceeding $3.17 U_p$ we set $d_2$ = 0.1. This helps us to reduce numerical noises stemming from the fact that the electronic continuum is discretized and the energy levels may not be evenly distributed. A similar modification has been proposed by Coccia \textit{et al.} \cite{coccia2016a,coccia2019}, who, however, attributed different values of $d$ to different CI eigenstates instead of different virtual orbitals.

Within the above-presented framework, we perform RT-TDCI calculations: first RT-TDCIS and standard RT-TDCISD, and then RT-TDCISD in reduced propagation space. For the truncation of the virtual-orbital space we use values of $\lambda$ = 1, 2, 3, 4 and 5. Similarly, for the truncation of the CISD state space we use values of $\eta$ = 1, 2, 3, 4 and 5. Finally, when selecting the CISD states based on squared partial norms of their vectors we use the thresholds for $N_k^{\mathrm{RS}}$ and $N_k^{\mathrm{RSBC}}$ equal to 0.5, 0.1, 10\textsuperscript{-2} and 10\textsuperscript{-3}.

\section{Results and discussion}
Let us start the analysis of the obtained results with a comparison between HHG spectra calculated at the RT-TDCIS and standard RT-TDCISD level of theory.
This will allow us to gauge the effects in the laser-driven electron dynamics stemming from full inclusion of dynamic correlation, as well as establish a reference for the results of propagations in the truncated CI space.
In each of the two Gaussian basis sets we obtain one occupied MO and 141 virtual MOs, so the full diagonalizations of the Hamiltonian matrices provide 142 CIS and 10153 CISD eigenstates.
The spectra obtained from real-time propagations in such complete CI spaces are presented on Figure 1.
We can notice is that at both levels of theory and at both laser intensities the atomic-augmented basis set (aug-cc-pVTZ+ARO3a for short) is able to correctly reproduce the spectrum profile described by the three-step model: a low energy perturbative region with a rapid decrease of harmonic intensity, the plateau region with nearly constant intensity of consecutive harmonic peaks, and the cutoff at energy close to $E_{\mathrm{cut}}$, after which the spectrum abruptly drops by several orders of magnitude.
On the other hand, the midbond-augmented basis set (aug-cc-pVTZ+ARO6m for short) fails completely to reproduce the 3SM predictions, with the harmonic plateau extending far beyond $E_{\mathrm{cut}}$ and a linear decrease of the peak intensities instead of a sharp, well-located cutoff.
This may seem surprising, especially since we use a very small value of escape length for orbitals which according to the single-electron picture should not contribute to harmonic generation, so this kind of artifacts should be largely reduced.
Moreover, the distributions of the CI eigenstates obtained in both basis sets (shown at Figure 2) are rather similar to each other, with the aug-cc-pVTZ+ARO6m eigenspectrum being only slighty shifted towards higher energies.

\begin{figure}
\centering
\includegraphics[width=0.8\linewidth]{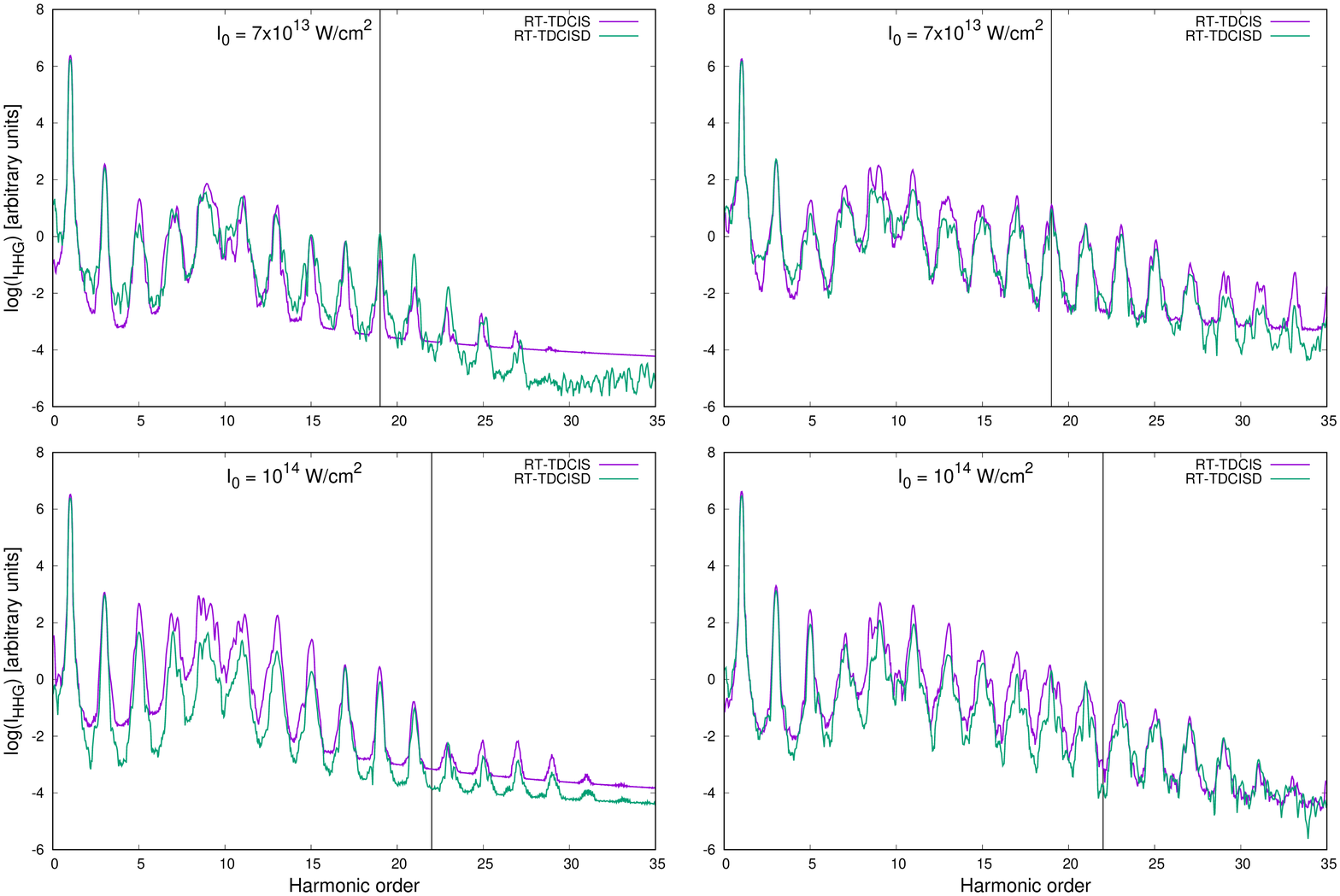}
\caption{HHG spectra of the H\textsubscript{2} molecule computed using the atomic-augmented basis set (left column) and the midbond-augmented basis set (right column). Solid vertical lines denote the cutoff positions predicted by the three-step model}
\end{figure}

\begin{figure}
\centering
\includegraphics[width=0.8\linewidth]{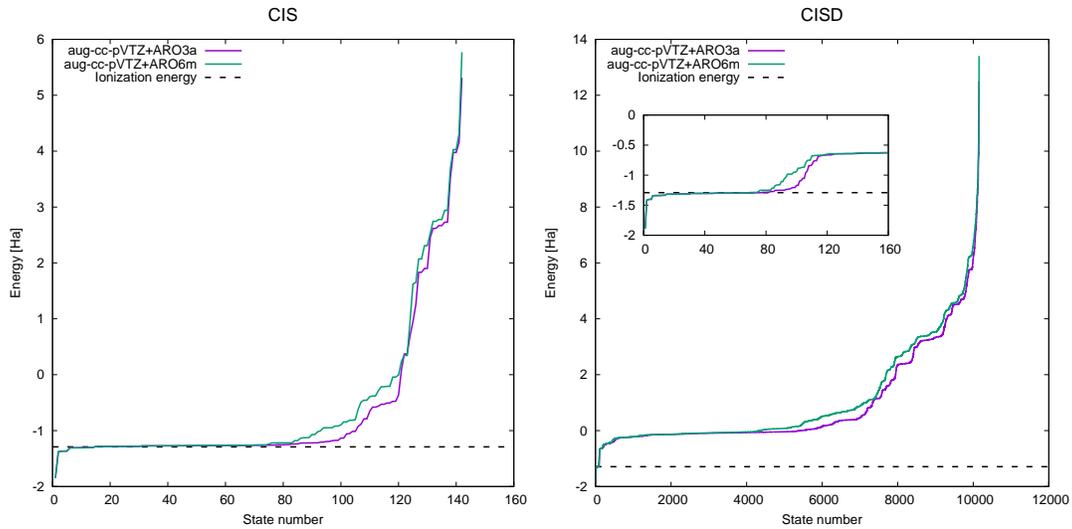}
\caption{Distributions of electronic energies of the H\textsubscript{2} molecule calculated using Gaussian basis sets with atomic augmentation and with midbond augmentation, at the CIS and CISD level of theory. Inset on the right plot shows the distribution of lowest 160 CISD states for an easier comparison with CIS results}
\end{figure}

We argue that the reason of the differences in spectra is that introducing the midbond augmentation inadvertently divides the CI eigenstates into two subgroups.
The bound and lower energy continuum states are described mostly with Dunning orbitals, while the description of higher energy continuum states is dominated by the contribution of ARO functions.
This may provide a reasonable picture in terms of electronic energies, but due to different shapes of molecular orbitals constructed from atomic and midbond functions, other observables of the eigenstates, such as the dipole moment expectaction values, may be reproduced with different accuracy between two subgroups.
When the wavefunction is excited from lower to higher energy part of energy spectrum, it experiences unphysical ``switches'' in the total dipole moment, that manifest in the HHG spectra in the form of additional harmonics beyond the theoretical cutoff positons, as well as more jagged shapes of the harmonic peaks.
Is is therefore clear that the midbond-augmented basis set is, at least in our case, not a reliable choice for modeling HHG spectra, and so we perform all further calculations only in the aug-cc-pVTZ+ARO3a basis set.

Aside from the performance differences between Gaussian basis sets, in all plots on Figure 1 one may notice some distinctive features of the spectra that can be attributed to the inclusion of double excitations.
For example, the RT-TDCISD spectra are characterized by lower HHG background intensity, with a more pronounced cutoff region (this obviously can be seen more clearly for the atomic-augmented basis set).
Such high difference in the intensity between the plateau peaks and of the baseline usually indicates a better quality of the theoretical spectra and a lower level of numerical noises.
Moreover, when using RT-TDCISD, the harmonic emission is decreased in the middle of the plateau, overall resulting in a more constant intensity of consecutive peaks, a feature we already reported in analogous calculations for the helium atom \cite{wozniak2022}.
In particular, lowering of peak intensities in the region close the ionization threshold (which in case of the H\textsubscript{2} molecule is located around 10th harmonic, corresponding to the ionization potential of 15.4 eV) is a clear instance of dynamic correlation in HHG.
The correlation effects are expected to be most significant in transitions to Rydberg or low energy continuum states, because the two electrons remain relatively close to each other and have to correlate their movements to minimize Coulombic repulsion.
Therefore, during deexcitation from a low energy state in the RT-TDCISD picture, the electron in the H\textsubscript{2}\textsuperscript{+} ion has its position already adjusted to the trajectory of the returning electron, so that the latter emits less \textit{bremsstrahlung}-like radiation than it would in the mean-field approximation of RT-TDCIS.
Naturally, the same effect is expected to occur in the perturbative region, but due to much higher intensity of first few harmonic peaks its practically unnoticeable in the logarithmic forms of the spectra.
By analogy, in transitions to high energy continuum states the ionized electron travels far away from the residual ion, so the electronic interaction weakens.
Thus, the recollision picture provided by RT-TDCIS and by RT-TDCISD is more similar, as are the intensities of harmonics at the end of the plateau.

Let us now move to the analysis of the simulations performed using different truncation algorithms, starting with the truncation of the virtual-orbital space.
The comparison of the HHG spectra computed using standard RT-TDCISD and using different values of $\lambda$ is presented on Figure 3, while the numbers of orbitals and CISD eigenstates used in real-time propagations are given in Table 1.
At first sight one can notice that the number of truncated virtual orbitals has a strong effect on the shapes of the spectra, with all curves varying in terms of peak shapes and intensities and the position of the baseline.
However, these changes are by no means systematic, i.e. the profiles of the spectra do not converge to the standard RT-TDCIS spectrum together with increasing value of $\lambda$.
For instance, at both laser conditions the lowest background intensity is obtained with $\lambda$=5, while the background intensity of RT-TDCIS is located between these of $\lambda$=3 and $\lambda$=4.
Another observation is that the virtual-orbital truncation scheme performs differently at different laser intensites, especially for lowest $\lambda$.
Spectra computed with $\lambda$=1 and $\lambda$=2 much closer resemble the standard RT-TDCISD at intensity $7 \times 10^{13}$ W/cm\textsuperscript{2} than at intensity $10^{13}$ W/cm\textsuperscript{2}.
Therefore, due to largely unpredictable effects of the truncation of virtual-orbital space on the obtained time-resolved observables, we consider this algorithm not a reliable choice for accelerating RT-TDCISD calculations.

\begin{figure}[h]
\centering
\includegraphics[width=0.8\linewidth]{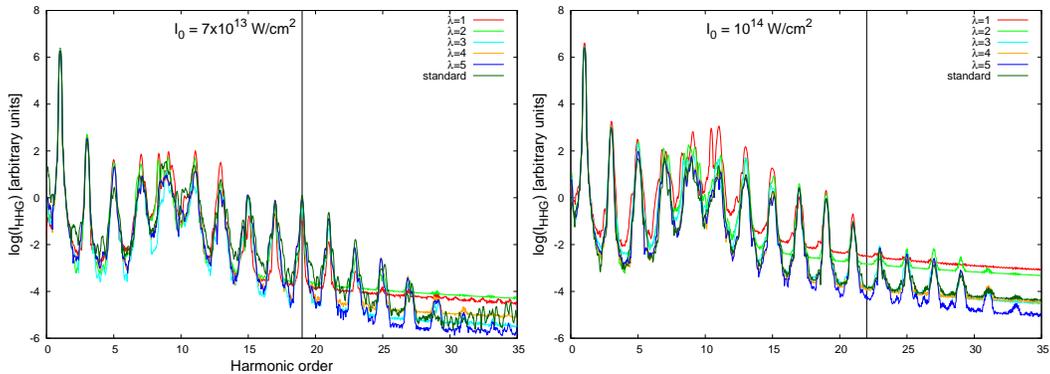}
\caption{HHG spectra of the H\textsubscript{2} molecule computed using standard RT-TDCISD and using RT-TDCISD with the algorithm of truncating the virtual-orbital space, for different values of $\lambda$}
\end{figure}

\begin{table}
\centering
\caption{Properties of the RT-TDCISD propagation spaces obtained using the algorithm of truncating the virtual-orbital space, for different values of $\lambda$ ($n_{vir}$ - number of selected virtual orbitals, $n_{\textrm{CISD}}$ - number of obtained CISD states)}
\begin{tabular}{ccccccc}
\toprule
\multicolumn{1}{c}{} & \multicolumn{3}{c}{$I_0$ = $7 \times 10^{13}$ W/cm\textsuperscript{2}} & \multicolumn{3}{c}{$I_0$ = $10^{14}$ W/cm\textsuperscript{2}} \\
\cmidrule(rl){2-4} \cmidrule(rl){5-7}
\textbf{$\lambda$} & {$\epsilon_{\mathrm{thr}}$ [Ha]} & {$n_{vir}$} & {$n_{\textrm{CISD}}$} & {$\epsilon_{\mathrm{thr}}$[Ha]} & {$n_{vir}$} & {$n_{\textrm{CISD}}$} \\
\midrule
1 & 0.4873 & 104 & 5565 & 0.6962 & 106 & 5778 \\
2 & 0.9746 & 109 & 6105 & 1.3923 & 119 & 7260 \\
3 & 1.4619 & 119 & 7260 & 2.0885 & 122 & 7626 \\
4 & 1.9492 & 120 & 7381 & 2.7846 & 124 & 7875 \\
5 & 2.4366 & 123 & 7750 & 3.4808 & 125 & 8001 \\
\bottomrule
\end{tabular}
\end{table}

A far more consistent picture emerges from the calculations employing the truncation of CISD eigenspace, as presented on Figure 4.
We observe some barely distinguishable differences between the spectra obtained with $\eta$=1 and the standard RT-TDCISD spectra, while the spectra obtained for $\eta$=2, 3, 4 and 5 completely overlap with the standard ones.
What is extremely important, this convergence rate is independent from the applied laser intensity, even though the number of CI eigenstates discarded using a particular value of $\eta$ may strongly depend on $I_0$ (Table 2).
This is an irrefutable proof that the states with energies above $E_{\mathrm{cut}}$ have little to no contribution to the laser-driven electron dynamics, at least from the HHG perspective.
If it were otherwise, we would observe much more distinct differences between the curves for not one but two reasons.
First, lack of sufficiently high energy states would force the wavefunction to populate low energy part of the eigenspectrum, which would translate into changes in the time-resolved dipole moment.
Second, in the employed ionization model states with high $E_k^{\textrm{CISD}}$ are expected to have very large ionization rates, due to dominant contribution of virtual orbitals with high $\gamma_a$.
Therefore, if the states beyond $E_{\mathrm{cut}}$ were relevant for the time-propagation, removing them would cause unphysical reflections of the wavefunction, further affecting the observables.
From the computational point of view, this algorithm seems to be highly efficient in accelerating RT-TDCISD calculations. 
By setting $\eta$=1, we can already capture most of the correlation effects in HHG offered by standard RT-TDCISD, even though the amount of states used in the actual real-time propagation is comparable to the amount of CIS states that can be obtained in the same basis set.
Nevertheless, for the optimal quality of results we recommend using a value of $\eta$=2, that can provide a complete convergence of the observables while still keeping the propagation space relatively compact.

\begin{figure}[h]
\centering
\includegraphics[width=1.0\linewidth]{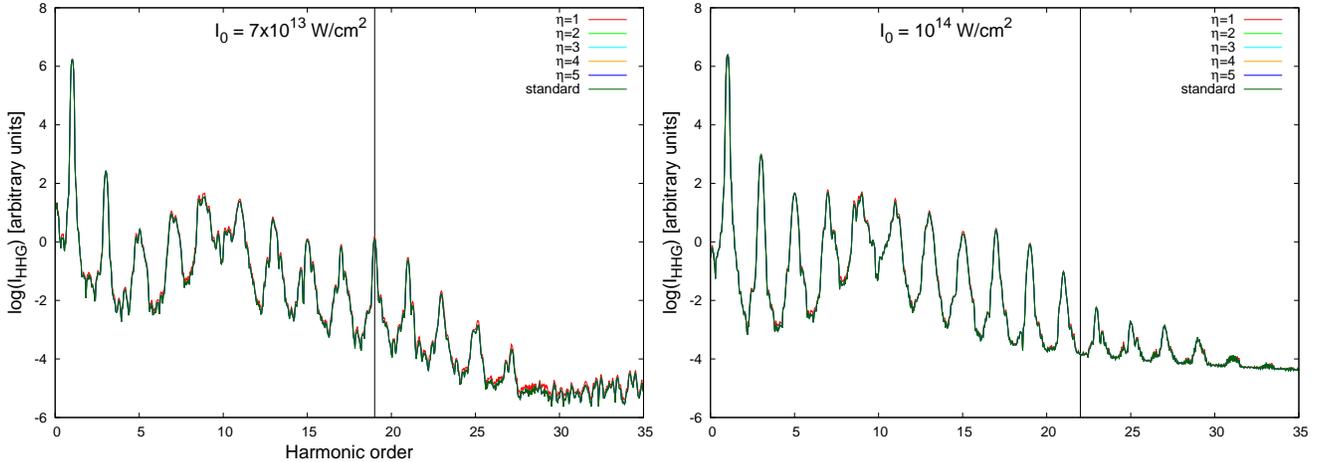}
\caption{HHG spectra of the H\textsubscript{2} molecule computed using standard RT-TDCISD and using RT-TDCISD with the algorithm of truncating the CISD eigenspace, for different values of $\eta$}
\end{figure}

\begin{table}[h]
\centering
\caption{Properties of the RT-TDCISD propagation spaces obtained using the algorithm of truncating the CISD eigenspace, for different values of $\eta$ ($n_{\textrm{CISD}}$ - number of selected CISD states)} 
\begin{tabular}{ccccc}
\toprule
\multicolumn{1}{c}{} & \multicolumn{2}{c}{$I_0$ = $7 \times 10^{13}$ W/cm\textsuperscript{2}} & \multicolumn{2}{c}{$I_0$ = $10^{14}$ W/cm\textsuperscript{2}} \\
\cmidrule(rl){2-3} \cmidrule(rl){4-5}
\textbf{$\eta$} & {$E_{\mathrm{thr}}$ [Ha]} & {$n_{\textrm{CISD}}$} & {$E_{\mathrm{thr}}$[Ha]} & {$n_{\textrm{CISD}}$} \\
\midrule
1 & 1.0817 & 110 & 1.2905 & 200 \\
2 & 1.5690 & 528 & 1.9867 & 5901 \\
3 & 2.0563 & 6031 & 2.6829 & 7210 \\
4 & 2.5436 & 7164 & 3.3790 & 7677 \\
5 & 3.0309 & 7467 & 4.0752 & 7962 \\
\bottomrule
\end{tabular}
\end{table}

Finally, on Figure 5 we present the comparison between the HHG spectra computed using RT-TDCISD with the selection of states based on the values of $N_k^{\mathrm{RS}}$ and $N_k^{\mathrm{RSBC}}$.
For both variants of this algorithm we see an even better convergence of the spectra profiles than we observed with the previous algorithm.
In fact, setting the threshold for either squared partial norm equal to 0.5 already results with a spectrum \textit{de facto} identical with the standard RT-TDCISD one.
Interestingly, if we look at the dimensions of the propagation spaces (Table 3), we notice that for the smallest threshold values the numbers of selected CISD states strongly differ depending on whether we choose them based on $N_k^{\mathrm{RS}}$ or on $N_k^{\mathrm{RSBC}}$.
However, along with increasing the threshold, this difference starts to decrease, and for the largest value of 0.5 in both selection schemes we arrive at the same number of 129 states.
After a careful analysis, we found these states to constitute exactly the same subset of the full CISD eigenspectrum.
This discards our hypothesis that the bound+continuum double excitations have a key role in laser-driven dynamics and should be treated on an equal footing with single excitations when performing correlated real-time calculations.
Instead, of all 10153 CISD eigenstates we have identified a subset of 129 ones which are dominated by the reference and/or singly excited configurations, and which are sufficient to describe the correlated laser-driven electron dynamics at both examined laser intensities.
The fact that we can reduce the dimension of the propagation space by two orders of magnitude while still being able to retrieve a fully correlated picture of HHG shows the true strength of this algorithm.

\begin{figure}[h]
\centering
\includegraphics[width=0.8\linewidth]{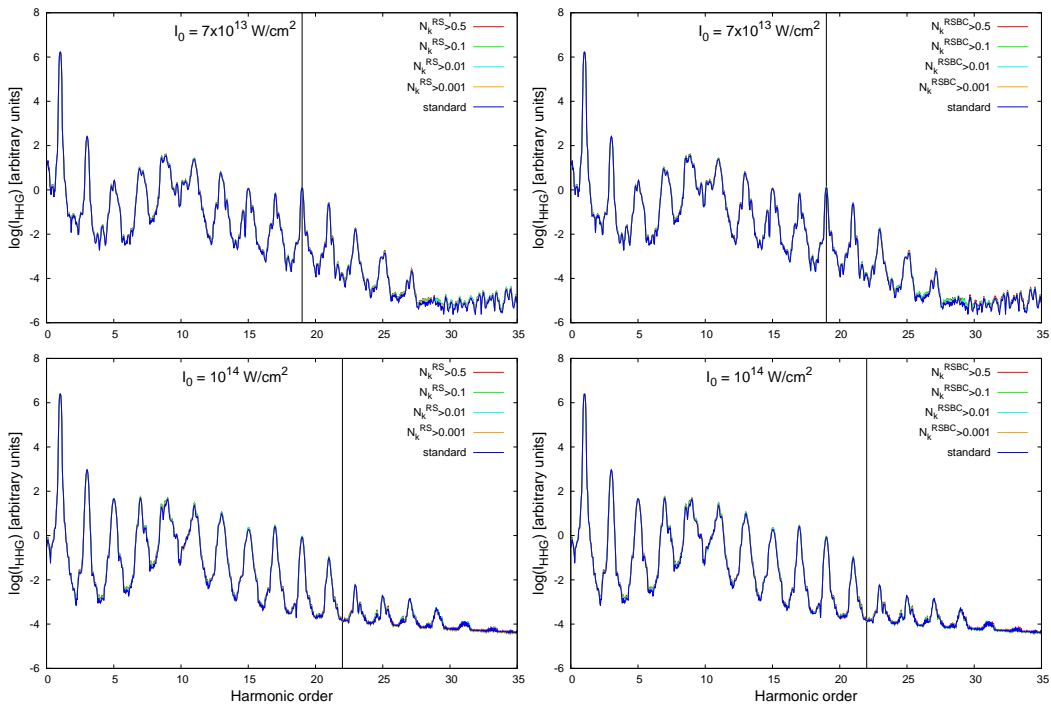}
\caption{HHG spectra of the H\textsubscript{2} molecule computed using standard RT-TDCISD and using RT-TDCISD with the algorithm of selecting CISD eigenstates based on the value of $N_k^{\mathrm{RS}}$ (left column) and $N_k^{\mathrm{RSBC}}$ (right column)}
\end{figure}

\begin{table}[h]
\centering
\caption{Dimensions of RT-TDCISD propagation spaces obtained using the algorithm of selecting CISD states based on their $N_k^{\mathrm{RS}}$ and $N_k^{\mathrm{RSBC}}$ values} 
\begin{tabular}{ccc}
\toprule
{partial norm} & {CISD states with $N_k^{\mathrm{RS}}$} & {CISD states with $N_k^{\mathrm{RSBC}}$} \\
{threshold} & {above threshold} & {above threshold} \\
\midrule
0.001 & 1382 & 5034 \\
0.01 & 492 & 2769 \\
0.1 & 173 & 529 \\
0.5 & 129 & 129 \\
\bottomrule
\end{tabular}
\end{table}

\section{Conclusion}
In this work we introduced, implemented, and tested several algorithms designed for reducing computational costs of RT-TDCISD simulations of laser-driven electron dynamics, with particular emphasis on modeling the high-harmonic generation process.
Their common feature is that they all rely on decreasing the number of CISD eigenstates the time-dependent CI vector is expanded into.
In the first one we limit the number of active virtual orbitals from which the configurations are constructed from, leaving only these with energies lower than a certain threshold linked to the HHG cutoff energy predicted by the three-step model.
In the second one, we apply a similar procedure, but instead of truncating the number of virtual orbitals, we compute the CISD eigenspectrum in a full MO space and then discard eigenstates above a predefined energy threshold.
Finally, in the third one, from the full CISD eigenspectrum we select the eigenstates to be included in real-time propagation depending on the contribution of certain types of excited configurations.

After performing test calculations of the HHG spectra of the hydrogen molecule, we found the latter two algorithms to be extremely efficient at selecting eigenstates that are of the most importance for the laser-driven electron dynamics.
At two examined laser intensities both of them were able to reduce the dimension of the propagation space by two orders of magnitude without affecting the time-resolved observables.
Moreover, both algorithms are largely independent from each other because they rely on different theoretical assumptions: that not all excited states are energetically accessible in a certain laser field, and that the matter-light interaction in the tunneling ionization regime of HHG is dominated by single-electron transitions, respectively.
Thanks to this, one can potentially combine both selection schemes to obtain even better results.

Aside for making the RT-TDCISD calculations much more affordable for larger systems and longer simulation times, these two algorithms may also be used as a tool for obtaining valuable information about the evolution of the wavefunction in the electromagnetic field.
For instance, we found out that at intensities typical for strong field experiments the HHG process in the H\textsubscript{2} molecule is driven mostly by a surprisingly small percentage of all excited CISD states, all of which are characterized by a dominant contribution of the ground and singly excited configurations and energies at most slightly exceeding the HHG energy cutoff of the three-step model.
As a consequence, we have confirmed that even if describing the laser-driven electron dynamics with a fully correlated \textit{ab initio} approach, the key predictions of 3SM still hold true.

On the other hand, we have shown that the method of truncating the active virtual-orbital space, routinely used in time-independent calculations of ground and low-excited states, is not a recommended approach in modeling strong field processes.
The excitations to very high-energy virtual orbitals usually have small but non-neglible contribution to all excited states and removing even a small portion of them may disrupt the description of HHG in a way that is hard to predict.

Finally, in this paper we also compare two approaches of supplementing the Gaussian basis sets with functions designed for the description of highly excited and continuum states: the one in which we place the additional functions on each atomic center and the one in which we place them at the center of the examined system. From our calculations, we conclude that for diatomic molecules the latter option, although initially appealing, produces less accurate results and should be used with caution.

\section*{Acknowledgements}

We acknowledge support from the National Science Centre, Poland (Symfonia Grant No.\\ 2016/20/W/ST4/00314).
The calculations have been partially carried out using resources provided by Wroclaw Centre for Networking and Supercomputing, grant No. 567.
M. L. also acknowledges support from: ERC AdG NOQIA; Ministerio de Ciencia y Innovation Agencia Estatal de Investigaciones (PGC2018-097027-B-I00/10.13039/501100011033, CEX2019-000910-S/10.13039/501100011033, Plan National FIDEUA PID2019-106901GB-I00, FPI, QUANTERA MAQS PCI2019-111828-2, QUANTERA DYNAMITE PCI2022-132919, Proyectos de I+D+I “Retos Colaboración” QUSPIN RTC2019-007196-7); MICIIN with funding from European Union NextGenerationEU(PRTR-C17.I1) and by Generalitat de Catalunya; Fundació Cellex; Fundació Mir-Puig; Generalitat de Catalunya (European Social Fund FEDER and CERCA program, AGAUR Grant No. 2017 SGR 134, QuantumCAT \textbackslash U16-011424, co-funded by ERDF Operational Program of Catalonia 2014-2020); Barcelona Supercomputing Center MareNostrum (FI-2022-1-0042); EU Horizon 2020 FET-OPEN OPTOlogic (Grant No 899794); EU Horizon Europe Program (Grant Agreement 101080086 — NeQST); ICFO Internal “QuantumGaudi” project; European Union’s Horizon 2020 research and innovation program under the Marie-Skłodowska-Curie grant agreement No 101029393 (STREDCH) and No 847648 (“La Caixa” Junior Leaders fellowships ID100010434: LCF/BQ/PI19/11690013, LCF/BQ/PI20/11760031, LCF/BQ/PR20/11770012, LCF/BQ/PR21/11840013). Views and opinions expressed in this work are, however, those of the author(s) only and do not necessarily reflect those of the European Union, European Climate, Infrastructure and Environment Executive Agency (CINEA), nor any other granting authority. Neither the European Union nor any granting authority can be held responsible for them.

\end{document}